\documentclass[journal,transmag]{IEEEtran}
\usepackage{cite}
\usepackage{amsmath}
\usepackage{amsfonts}
\usepackage{algorithmic}
\usepackage{array}
\usepackage{hyperref}
\usepackage{cancel}

\usepackage{fixltx2e}
\usepackage{graphicx}
\usepackage{url}

\newcommand{\ignore}[1]{}
\newcommand{\myfig}[1]{Fig. \ref{#1}}

\begin{document}
\title{On Universal Codes for Integers: Wallace Tree, Elias Omega and Variations}
\author{
\IEEEauthorblockN{Lloyd Allison, Arun Konagurthu and Daniel Schmidt}
\IEEEauthorblockA{Faculty of Information Technology, Monash University, Clayton, Victoria 3800, Australia}
\thanks{Corresponding author: Lloyd Allison (lloyd.allison@monash.edu).}}

\IEEEtitleabstractindextext{%
\begin{abstract}
A universal code for the (positive) integers can be used
to store or compress a sequence of integers.
Every universal code implies a probability distribution on integers.
This implied distribution may be a reasonable choice
when the true distribution of a source of integers is unknown.
Wallace Tree Code (WTC) is a universal code for integers based on binary trees.
We give the encoding and decoding routines for WTC and
analyse the properties of the code
in comparison to two well-known codes, the Fibonacci and Elias omega codes.
Some improvements on the Elias omega code are also described and examined.
\end{abstract}

\begin{IEEEkeywords}
    Universal integer codes, variable length code, prefix codes
\end{IEEEkeywords}}

\maketitle

\IEEEdisplaynontitleabstractindextext

\section{Introduction}

\IEEEPARstart{U}{niversal} codes for the positive integers, $N\in \mathbb{Z}^+$,
are of interest for at least three reasons.
The first is in everyday data compression where such a code can
be used to store or to transmit a sequence of integers when
their true probability distribution is not known.
The second is in inductive inference, where
a countable set of hypotheses in a statistics and
machine learning task is mapped to the set of positive integers:
If the true distribution of the set of hypotheses in an
inference problem is unknown
but they can be plausibly ordered in non-increasing probability,
then the $i^{\text{th}}$ hypothesis can be assigned the probability $2^{-|w(i)|}$,
where $w(i)$ is the code-word of integer $i$, whose length is represented as $|w(i)|$.
Finally, it must also be admitted that there is simply a fascination in
trying to devise an efficient code for truly enormous integers.

Elias~\cite{Eli75} defined a code having the \textit{universal} property as
one where
the code-word length is monotonically increasing and
``assigning messages in order of decreasing probability to codewords in
order of increasing length gives an average code-word length,
for any message set with positive entropy,
less than a constant times the optimal average codeword length for
that source.''
If the source has distribution $\Pr(\cdot)$ and entropy
$$
H = \sum_{\forall N > 0} \Pr(N)\log\left(\frac{1}{\Pr(N)}\right)
$$
then, for any universal code $w(\cdot)$ for positive integers, 
$$
E_{w} = \sum_{\forall N > 0} \Pr(N) |w(N)| < K \cdot H,
$$
where $K$ is a constant \textit{independent} of $\Pr(\cdot)$.
The latter sum is at least finite, although the distribution implied by
a universal code must itself have infinite entropy.
Naturally we hope that $K$ is not large.
Elias also defined an \textit{asymptotically optimal} code as one where
the ratio
$$
\frac{E_{w}}{\max(1, H)} \le R(H) \le K,
$$
where $R$ is a function of $H$ 
with
$
\lim_{H\rightarrow \infty} R(H) = 1.
$

Wallace proposed a universal code for integers \cite{Wal05} (WTC)
inspired by binary trees.
He suggested that its implied probability distribution is a reasonable choice
to use when the true distribution of a source of integers is unknown.
In the following sections,
the WTC, Fibonacci and Elias omega codes are summarised.
Encoding/Decoding routines and asymptotic analysis are given for WTC.
The properties of the three codes,
particularly the lengths of code-words, are compared. 
Ways of improving the Elias omega code for large integers are
also discussed. 

Note, if we have non-negative integers, a code for $N \ge 1$
can be shifted and used for $N \ge 0$ by adding one before
encoding and subtracting one after decoding.
If we have \textit{all} of the integers to deal with,
they can be ordered as $[0,1,-1,2,-2,3,\ldots]$, say, and
$N$ can be encoded according to its position in the list.
Encoding and decoding routines for the codes described
can be found, and interactively experimented with, at
\href{http://www.allisons.org/ll/MML/Discrete/Universal/}{www.allisons.org/ll/MML/Discrete/Universal/}.

\section{Three universal codes}

The Fibonacci and Elias omega codes for positive integers are described below
for later comparison to the Wallace tree code (WTC) for positive integers.
The main focus is in the Elias omega code and WTC;
the Fibonacci code is included as a fixed point of comparison to
the other codes.

\subsection{Fibonacci code for integers}
\label{sec:Fibonacci}

The Fibonacci code \cite{AF87} is based on the Fibonacci numbers
$F_0=1, F_1=1, \text{~and~} F_j=F_{j-1}+F_{j-2}, \text{~for~} j\ge2$,
which are $[1,1,2,3,5,8,\ldots]$.
The code ignores $F_0$ and uses $F_1$ and upwards.
To encode an integer $N \ge 1$ do as follows:
\begin{itemize}
    \item Find the largest $F_{j_1}$ that is less than or equal to $N$ and
remember $j_1$.
\item Repeat for $N'=N-F_{j_1}$, finding $j_2, j_3$, and so on until zero remains.
\end{itemize}
We have $N=F_{j_1}+F_{j_2}+\ldots$.
The code-word for $N$ consists of $j_1+1$ bits.
If $F_k$ appears in the sum for $N$ set the $k^\text{th}$ bit of
the code-word to `1' otherwise set it to `0'.
Set the last bit to `1'.
It is easy to see that the encoding is well defined and
that each $j_k < j_{k-1}-1$.
Note that ``11'' cannot appear in the code-word except at
the very end and that a code-word of length $|w(N)|$ contains at
least two and up to $2+\lfloor(\frac{|w(N)|-2)}{2}\rfloor$ `1's.

\subsection{Elias omega ($\omega$) code for integers}
\label{sec:Elias}

We first introduce the following variation on the Elias omega code that gives
identical code-word lengths to the original definition \cite{Eli75} and
differs in only minor details.
The code-word for an integer $N \ge 1$ consists of
one or more sections: zero or more \textit{length} sections followed
by one \textit{value} section.
The value section is simply the usual binary representation of $N$
in $\lfloor \log_2(N) \rfloor + 1$ bits;
note that the value section starts with a `1'.

The code-word for $N=1$ is simply ``1'';
it is the only code-word that starts with a `1'.
The code-word for $N \ge 2$ has at least one length section before
the final value section.
In general the value section by itself is not sufficient because
a decoder does not know how long it is when $N \ge 2$.
The solution is to first encode the length of the
value section minus one (the length must be $\ge 2$ when $N \ge 2)$,
recursively, until the length$-1$ of a length$-1$ of $\ldots$ of a length$-1$
gives one.

The leading bit of each section would, on the face of it,
be a `1' so that position can instead be used as a flag to
indicate either a length section (`0') or the
final value section (`1').
The decoder notes the flag.
In the case of a `0' it then switches it to a `1' before
computing the length of the next section.
If present ($N \ge 2$) the first length section is
just ``0'' which stands for one.
If $N \ge 4$ the second length section is either ``00''
which stands for two or ``01'' which stands for three,
and so on.

Note that the omega code is very similar to, and can be thought of as
an optimized and shifted version of,
the Levenstein code \cite{Lev68} which is defined for $N \ge 0$.
Also note that Rissanen \cite{Ris83} defined the $\log^*$ code as
an approximation to the omega code;
$\log^*(N) = c + \log_2(N) + \log_2(\log_2(N))
  + \log_2(\log_2(\log_2(N)))  \ldots$,
all positive terms, where $c$ is a normalising constant,
and $\Pr(N) = 2^{-\log^*(N)}$.

\begin{figure}
\begin{verbatim}
function omega_r_enc(t_enc)
 { function enc(N)  // bigInt N >= 1
    { var todo=N, nSect, nTet, CW="";
      for( nTet = 1; ; nTet ++ )
       { for( nSect = 1; ; nSect ++ )
          { var section = todo.binary();
            var len = section.length;
            if( len == 1 ) break;
            // else trim section
            section =
               section.substring(1,len);
            CW = section + CW;
            todo = N.fromInt(len-1);
          }//for nSect
         if( nSect == 1 ) break
         todo = N.fromInt(nSect-1);
       }//for nTet
      CW = t_enc(N.fromInt(nTet)) + CW;
      return CW;
    }//enc(N)
   return enc;
 }//omega_r_enc(t)

// e.g. ...
function omega_star_enc(N)
   = omega_r_enc(omega_enc)(N);
\end{verbatim}
    \caption{Encoder for omega\textsubscript{r}(t)(N) in JavaScript-styled pseudocode.}
\label{fig:star}
\end{figure}

\subsubsection{The omega variations}
\label{sec:variations}

The Elias omega code in effect uses a unary code
(``0\ldots'' $\Rightarrow$ length section,
``1\ldots'' $\Rightarrow$ final value section)
to indicate the number ($\ge 1$) of sections in a code-word.
Elias chose the name omega for the code because
he considered it to be ``penultimate'', that is ``not quite ultimate''
(p.200)\cite{Eli75}.
(That being so, the name \textit{psi}, say,
would have left some room to name codes that are closer to the ultimate.)
He noted that the unary code could
be replaced with his gamma, delta or omega codes.

In fact the leading bits of the sections
can be moved to the front of the code-word and
the unary code can be replaced by \textit{any} other code
(universal or not) for positive integers -- say Fibonacci or WTC.

Define omega\textsubscript{p}(s) to be
the Elias omega code modified and \textit{parameterised} to use
an integer code `s' for the number of sections.
The code-word for $1$ is ``1''.
For $N > 1,$
the code-word is the omega code-word for $N$
with the leading bit of every section trimmed away, and the result
preceded by the `s' code-word for the number of sections.
omega\textsubscript{p}(unary) is equivalent to the usual omega code.
The code omega\textsubscript{p}(Fibonacci) would make
code-words of one, two and three sections longer and
those of six or more sections shorter
than under omega.
Let omega\textsuperscript{2} = omega\textsubscript{p}(omega)
which uses the omega code for the number of sections.
This code would make
code-words of two, four and five sections longer and
those of seven or more sections shorter
than the usual Elias omega code.

Even omega\textsuperscript{2} is not ultimate.
It is possible to define a code, omega\textsubscript{r}(t),
that uses itself, recursively, to state the
number of sections (minus one) in an omega code-word.
Note that omega\textsubscript{r}($\cdot$) needs some other integer code 
to encode the number of \textit{tetrations}, i.e.  the number of times
that omega\textsubscript{r} is applied.
Let omega\textsuperscript{*} = omega\textsubscript{r}(omega).
For example, $N=36$ is encoded using  omega\textsuperscript{*} as follows (spaces added in code-words below for readability):
\begin{eqnarray*}
    N = 36 \implies 
    \text{trim}(\text{omega}(36)) 
    &=& {\cancel{0}~ \cancel{0}0~ \cancel{0}01~ \cancel{1}00100}\\
    &=& 0~ 01~ 00100\\
    \text{\#sections} = 3 \implies 
    \text{trim}(\text{omega}(3)) 
    &=& {\cancel{0}~ \cancel{1}1}\\
    &=& 1\\
    \text{\#sections} = 1 \implies 
    \text{trim}(\text{omega}(1)) 
    &=& \cancel{1}\\
    &~&\text{Stop.}\\
    \text{\# tetrations} = 3.\\
    \text{Encoded using omega(3)} &=& 0~11\\
    \implies \text{omega}^*(36) = 0 11 1 0 01 00100 
\end{eqnarray*}
where $\text{trim}(.)$ removes the
leading bit of each length and value section of recursively applied omega code-words.
\ignore{
}

Comparing against the omega code, omega\textsuperscript{*} would make
code-words  of integers with nine or more sections shorter
than under omega and
the smallest corresponding integer is $2\uparrow\uparrow8$
in Knuth's up-arrow notation.
%
%
An encoder for omega\textsubscript{r}
is given in ~\myfig{fig:star}.

Even omega\textsuperscript{*} is not ultimate
because one can reconsider the encoding of
the number of tetrations but the integers
for which there would be any further
improvement would be ``very large'' indeed.

\subsection{Wallace tree code (WTC) for integers}
\label{sec:WTC}

The Wallace tree code for integers \cite{Wal05}\cite{All18} are based on full binary trees, and hence depend on the Catalan numbers $C_f, \forall f \ge 0$,  with the first few numbers being $C_0=1, C_1=1, C_2=2, C_3=5, C_4=14,$ and so on.

Any full binary tree consists of $f \ge 0$ \textit{fork} (i.e. internal) nodes and
$f+1$ \textit{leaf} nodes.
Each fork node has \emph{exactly} two sub-trees and
each leaf node has \emph{zero} sub-trees.

The number of full binary trees containing $f$ fork nodes is
the $f^{\text{th}}$ Catalan number, defined as:
$$C_f = \frac{1}{f+1} {2f \choose f} = \frac{(2f)!}{(f+1)!f!}.$$
This yields the recurrence $C_{f+1}=\frac{2(2f+1)C_f}{f+2}$. Furthermore,
$\lim_{f\rightarrow\infty}\frac{C_{f+1}}{C_f} = 4,$ and
 $C_{f+1} = \sum_{j=0}^{f} (C_j\cdot C_{f-j})$ \cite{Seg61}.
Finally, it is  useful to define the \emph{cumulative} Catalan numbers
$cC_f = \sum_{j=0}^f C_f, \forall f \ge 0$,
which are $cC_0=1, cC_1=2,cC_2=4, cC_3=9, cC_4=23,$ and so on.

\begin{figure}
\centering
\includegraphics[width=0.3\textwidth]{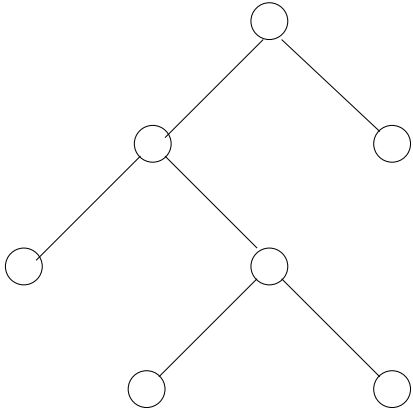}
\caption{Tree having code-word 1101000}
\label{fig:egTree}
\end{figure}

As a preliminary matter, note that a full binary tree
can be encoded \cite{WP93} during a prefix traversal of the tree,
outputting a `1' for each fork and a `0' for each leaf (e.g.
see ~\myfig{fig:egTree}).
The end of the tree's code-word is indicated upon
reaching one more `0' than `1's, and this event
does not happen earlier within the code-word so
this is a prefix code.
It is easy to recover the tree from the code-word.

\begin{figure}
\centering
\includegraphics[width=0.45\textwidth]{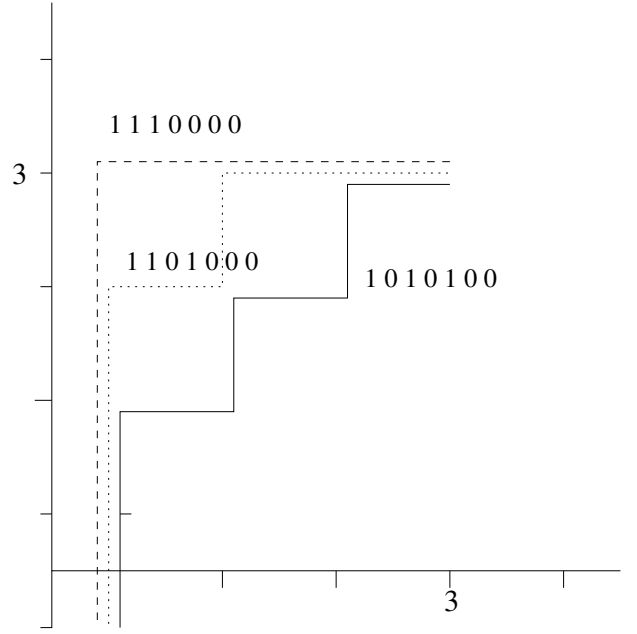}
\caption{Three Dyck paths.}
\label{fig:DyckPaths}
\end{figure}

Reading a `1' as left and a `0' as down,
a tree's code-word can also be interpreted as
encoding a Dyck path \cite{Rus08} in a square lattice from
some, initially unknown, point on the diagonal $(f,f)$ to $(0,-1)$, that is,
a zig-zag path that does not go below the diagonal
until it terminates with the final `0'.
\myfig{fig:DyckPaths} shows three of the five paths
from $(3,3)$ to $(0,-1)$ with their code-words
including the lexicographically
first `1010100' and
last  `1110000'.
The number of paths (\myfig{fig:paths})
from row $r$, column $c$, to $(0,0)$ is given by
\begin{equation}
\begin{split}
    \mathrm{paths}_{r,c} & = 0, \text{~if~} c > r,  \\
    \mathrm{paths}_{0,0} & = 1,    \\
    \mathrm{paths}_{r,0} & = 1, r \ge 0,  \\
    \mathrm{paths}_{r,c} & = \mathrm{paths}_{r-1,c} + \mathrm{paths}_{r,c-1}, \text{~otherwise}.
\end{split}
\end{equation}
It can be shown that $\mathrm{paths}_{f,f}=C_f$.

\begin{figure}
\centering
\begin{tabular}{ r|r r r r r r r r r  }
6 &   1  & 6  & 20  & 48  & 90  & 132  & 132  & 0   & \ldots \\
5 &   1  & 5  & 14  & 28  & 42  &  42  &   0  & \ldots \\
4 &   1  & 4  &  9  & 14  & 14  &   0  & \ldots \\
3 &   1  & 3  &  5  &  5  & 0   & \ldots \\
2 &   1  & 2  &  2  &  0  & \ldots \\
1 &   1  & 1  &  0  & \ldots \\
0 &   1  & 0  & \ldots   \\
\hline
  $r \uparrow$ &   0  & 1  &  2  &  3  &  4  &  5  &    6  & ... & $\rightarrow c$ \\
\end{tabular}
    \caption{$\mathrm{paths}_{r,c}$}
\label{fig:paths}
\end{figure}

The code for integers is most easily
explained for integers $N \ge 0$; call this version of the code WTC0.
The full binary trees are sorted on their code-word lengths and,
for a given length, lexicographically.
For a given length, the first code-word is
of the form $(10)^f0$ and the last $1^f0^{f+1}$.
Integer $N \ge 0$ is given the code-word of the $N^{th}$
full binary tree (in the lexicographic order of full binary trees, counting from zero).

Encoding and decoding routines are given in ~\myfig{fig:WTC}.
The code-word for $N=0$ is ``0''.
For $N > 0 $, the $C_f$ integers in the range $cC_{f-1} < N \le cC_f$,
all have code-words of length $2f+1$.
$f$~can be found by
searching for the largest cumulative Catalan number, $cC_{f-1}$,
that is less than $N$.
The code-word for $N$ is the $K^{th}$
of the lexicographically ordered code-words of length $2f+1$
where $K = (N - cC_{f-1})$.
The code-word can be found using $\mathrm{paths}_{r,c}$.
Starting with an empty code-word at position $(r,c)$ where $r=f, c=f$.
If $K > \mathrm{paths}_{r-1,c}$ then
append a `1' (move left, $c=c-1$) to the code-word and
we need a code-word at least that much further up
the rankings ($K=K-\mathrm{paths}_{r-1,c}$).
Otherwise, append a `0' (move down, $r=r-1$).
Repeat until $r=c=0$.

\begin{figure}
\begin{verbatim}
function WTC0enc(N)
 { if( N.isZero() ) return "0";
   var f=cCsearch(N); //min f st cC(f)>N
   var K=N.sub(cCatalan(f-1));
   var r=f, c=f, ans="";
   while( r > 0 )
    { var Decr = paths(r-1, c);
      if( K.GE( Decr ) )
           { ans = ans + "1"; c -- ;
             K = K.sub( Decr ); }
      else { ans = ans + "0"; r -- ; }
    }//while
   return ans + "0";
 }//WTC0enc

function WTC0dec(str)
 { // assumes str is a valid code-word
   if( str == "0" ) return Zero;
   var i, f = Math.floor(str.length/2);
   var r = f, c = f, Ans = cCatalan(f-1);
   for( i = 0; i < str.length; i ++ )
      if( str.charAt(i) == "0" ) r -- ;
      else /* "1" */
       { Ans = Ans.plus( paths(r-1, c) );
         c -- ;
       }
   return Ans;
 }//WTC0dec
\end{verbatim}
    \caption{Encoder and decoder for WTC0 for all integers $\ge 0$.}
\label{fig:WTC}
\end{figure}

The decoding routine (also in \myfig{fig:WTC}) follows
similar logic to the encoding routine.
A valid code-word, $str$, contains $f=\lfloor |str| / 2 \rfloor$ `1's
and $f+1$ `0's and no proper prefix contains more `0's than `1's.
As $str$ is processed from left to right,
every `1' (move left) means that $str$
is known to be at least $\mathrm{paths}_{r,c}$ further up in rank
amongst the code-words of this length.
Repeat until the end of the code-word.

As noted before, it is easy to ``shift''  WTC0 code (which encode integers $N\ge 0$) to instead encode integers $N \ge 1$.
Call the shifted code WTC1 if we need to distinguish between it and WTC0.

\subsection{Examples}
\label{sec:examples}

Table \ref{table:codewords} gives examples of
integers coded under the three codes.
Note that the lengths of Fibonacci code-words increase in
steps of one from time to time as $N$ grows. 
The lengths of WTC1 code-words increase in
steps of two.
Lengths in the Elias omega code increase in steps of
various sizes, for example increasing by four on going
from $N=15$ to $N=16$ when the value section grows
by one and a whole new length section is added;
there is no upper limit to the step size.

\begin{table}
\centering
\caption{Examples of code-words}
\label{table:codewords}
\begin{tabular}{ |l||l|l|l| }
\hline
N      &  Fib     &    Elias $\omega$  &    WTC1  \\
\hline\hline
1      &   11     &    1        &    0  \\
2      &   011    &    010      &    100  \\
3      &   0011   &    011      &    10100  \\
4      &   1011   &    000100   &    11000  \\
5      &   00011  &    000101   &    1010100  \\
6   & 10011  & 000110  & 1011000     \\
7   & 01011  & 000111  & 1100100     \\
8   & 000011  & 0011000  & 1101000     \\
9   & 100011  & 0011001  & 1110000     \\
10  & 010011  & 0011010  & 101010100     \\
11  & 001011  & 0011011  & 101011000     \\
12  & 101011  & 0011100  & 101100100     \\
13  & 0000011  & 0011101  & 101101000     \\
14  & 1000011  & 0011110  & 101110000     \\
15  & 0100011  & 0011111  & 110010100     \\
16  & 0010011  & 00000010000  & 110011000     \\
17  & 1010011  & 00000010001  & 110100100     \\
18  & 0001011  & 00000010010  & 110101000     \\
19  & 1001011  & 00000010011  & 110110000     \\
20  & 0101011  & 00000010100  & 111000100     \\
21  & 00000011  & 00000010101  & 111001000     \\
22  & 10000011  & 00000010110  & 111010000     \\
23  & 01000011  & 00000010111  & 111100000     \\
24  & 00100011  & 00000011000  & 10101010100     \\
100  & 00101000011  & 0000101100100  & 1011101001000     \\\hline\hline
\multicolumn{4}{|c|}{Below code-word \emph{lengths} are shown, instead of code-words}\\\hline\hline
$10^2$ & $|CW|=11$  & $|CW|=13$      & $|CW|=13$  \\
$10^3$ & $|CW|=16$  & $|CW|=17$      & $|CW|=17$  \\
$10^4$ & $|CW|=20$  & $|CW|=21$  & $|CW|=21$     \\
$10^5$ & $|CW|=25$  & $|CW|=28$  & $|CW|=25$     \\
$10^6$ & $|CW|=30$  & $|CW|=31$  & $|CW|=27$     \\
$10^7$ & $|CW|=35$  & $|CW|=35$  & $|CW|=31$     \\
$10^8$ & $|CW|=39$  & $|CW|=38$  & $|CW|=35$     \\
$10^9$ & $|CW|=44$  & $|CW|=41$  & $|CW|=39$     \\
googol  & $|CW|=480$  & $|CW|=349$  & $|CW|=345$     \\
\hline
\end{tabular} \\
    where googol$=10^{100}$, $|CW|$ = code-word length (in bits)\\
\end{table}

\section{Implied probability distributions}
\label{sec:implied}

An efficient code for the positive integers $N \ge 1$
implies a probability distribution on them in which
$\Pr(N) = 2^{-|w(N)|}$.
The Fibonacci, Elias omega and Wallace tree (WTC1) codes
all imply proper probability distributions on the
positive integers:
Consider an infinite string of bits generated at random,
independent and identically distributed (i.i.d.),
with $\Pr(\text{`0'}) = \Pr(\text{`1'}) = 0.5$.
For each code, the infinite string has some prefix,
of length $L$ and probability $\frac{1}{2^{L}}$, which is
a valid code-word in that code.
In principle, by removing the prefix and repeating forever,
all possible code-words will be sampled in proportion to
their probabilities under the corresponding distribution.

\subsection{Fibonacci:}

\begin{itemize}
\item There must be a first occurrence of ``11'' in
  the infinite string. It marks the end of a prefix of
  length $L \ge 2$ which is a valid code-word in the Fibonacci code.
        The probability of the prefix is $2^{-L}$.
  There are $i=L-2$ positions before the final ``11''.
  Each of these position can hold `0' or `1' but no two
  adjacent positions, other than the last, can both hold `1'.
\item There are $F_i$ code-words of length $L=i+2$, $i \ge 0$
  (a code-word of length $L$ can be formed by prepending one of
  length $L-1$ with a `0', or prepending one of length $L-2$ with a ``10'').
\item The total probability of those integers having
  code-words of length $(i+2)$ is $\frac{F_i}{2^{i+2}}, i \ge 0$.
\item The total probability of all positive integers,
  $\sum_{i\ge 0} \frac{F_i}{2^{i+2}}$, must be one.
\end{itemize}

\subsection{Elias omega:}

\begin{itemize}
\item An Elias code-word is made up of zero or
  more length sections followed by one value section.
  The infinite string of bits starts either `1' or `0'.
  If it starts `1', that itself is a prefix which is
  an Elias code-word for the value one, of probability $0.5$.
  If it starts `0' that is decoded as $1$
  (the lead bit having been changed) which indicates that
  a section of length $2=1+1$ follows.
  If the next section starts `1' it is a value.
  If it starts `0' it is a length, ``00'' giving
  $3=2+1$ or ``01'' giving $4=3+1$. And so on.
\item Eventually a section starting `1', of length $s$,
  will appear marking the end of a code-word of length $L$.
\item There are $2^{s-1}$ code-words of length $L$.
\end{itemize}

\subsection{WTC:}

\begin{itemize}
\item Because a one-dimensional random walk
  (`0' ~ left, `1' ~ right, say)
  returns to the origin with probability one,
  the infinite string has some prefix of length $L=(2f+1), i \ge 0$,
  that is a valid WTC1 code-word.
  The probability of the prefix is $\frac{1}{2^{2f+1}}$.
\item There are $C_f$ code-words of length $(2f+1)$.
\item The total probability of those integers having code-words
  of length $(2f+1)$ is $C_f/2^{2f+1}, f \ge 0$.
\item The total probability of all positive integers,
  $\sum_{f \ge 0} C_f / 2^{2f+1}$, must be one.
\end{itemize}

\vspace{4mm}

Table \ref{table:cummulative} gives cumulative probabilities
for integers having code-words upto a given length.
Note that an Elias omega code-word of $1,000,000$ bits needs at least
six sections, the maximum lengths of sections being
$1, 2, 4, 16, 65536, 2^{65536},\ldots$.
The probability of having five or fewer sections is
$1/2+1/4+1/8+1/16+1/32 = 1 - 0.03125 = 0.96875$,
slightly less than 0.9692.

\begin{table}
\centering
\caption{Cumulative probabilities up to code-word length $|w(N)|$.
    }
\label{table:cummulative}
\begin{tabular}{ |r||l|l|l| }
\hline
$|w(N)|$ (bits) & Fibonacci & Elias $\omega$ & WTC1 \\\hline
\hline
1     & 0     & 0.5   & 0.5    \\
2     & 0.25  & 0.5   & 0.5    \\
3     & 0.375 & 0.75  &  0.625    \\
4     & 0.5   & 0.75  & 0.625    \\
10    & 0.859 & 0.875 & 0.754    \\
100       & 0.999... & 0.947   & 0.920    \\
1000      & 0.999... & 0.957   & 0.975    \\
10000     & 0.999... & 0.963   & 0.992    \\
100000    & 0.999... & 0.9688  & 0.997    \\
1000000   & 0.999... & 0.9692  & 0.9992   \\
\hline
\end{tabular}  \\
\end{table}

\section{Comparative code-word lengths}
\label{sec:comparative}

For integers less than $100,000$ the ``lead'', in the sense of
having the shortest code-words, changes hands between the
three codes but is often held by the
Fibonacci code (table \ref{table:lead}) until $N=317,811$ where
it falls out of contention.
Beyond that, and
up to the $506$ decimal digit integer corresponding to $cC_{847}+1$,
WTC has the shortest code-words, rarely equalled by the
Elias omega code (i.e., for values
between $cC_{134}+1$ and $2^{255}-1$, both codes
using $269$ bits). To contextualize these numbers, this is past the size of the
human genome ($3.2 \times 10^9$ base-pairs),
the estimated number of baryons in the universe ($10^{80}$) and
one googol ($10^{100}$). 
Therefore beyond some further point the Elias omega code
must take the lead either permanently or at least most of the time --
see section~\ref{sec:asypmtotics}.

\begin{table}
\centering
    \caption{Code-word lengths (in bits) for varying $N$ (chosen to highlight early points of change)  across Fibonacci, Elias and WTC1. Shortest code-word lengths for a given $N$ are asterisked}
\label{table:lead}
\begin{tabular}{ |r||l|l|l| }
\hline
N & Fibonacci & Elias $\omega$ & WTC1 \\\hline
\hline
1     & 2    & 1*    & 1*   \\
2     & 3*   & 3*    & 3*   \\
3     & 4    & 3*    & 5   \\
4     & 4*   & 6     & 5   \\
13    & 7*   & 7*    & 9   \\
16    & 7*   & 11    & 9   \\
610   & 15*  & 17    & 15*   \\
627   & 15*  & 17    & 17   \\
1597  & 17*  & 18    & 17*   \\
2057  & 17*  & 19    & 19   \\
4181  & 19*  & 20    & 19*   \\
6765  & 20   & 20    & 19*   \\
6919  & 20*  & 20*   & 21   \\
8192  & 20*  & 21    & 21   \\
10946 & 21*  & 21*   & 21*   \\
16384 & 21*  & 22    & 21*   \\
17711 & 22   & 22    & 21*   \\
23715 & 22*  & 22*   & 23   \\
28657 & 23   & 22*   & 23   \\
32768 & 23*  & 23*   & 23*   \\
46368 & 24   & 23*   & 23*   \\
65536 & 24   & 28    & 23*   \\
82501 & 25*  & 28    & 25*   \\
\hline
\end{tabular}  \\
\end{table}

For each of the three codes, integers come in ``blocks'' that
contain integers having code-words of the same length under that code.
The sizes of the blocks differ between the codes.
For WTC, the blocks are
$[1,1], [2,2], [3,4], [5,9], ..., [cC_{f-1}+1,cC_f], ...$
having code-lengths $1, 3, 5, 7, ..., 2f+1,...$ bits respectively.
For $f \ge 848$ and up to at least $f=1000$,
$|\text{WTC}(cC_{f-1}+1)| = |\text{omega}(cC_{f-1}+1)|$ and
$|\text{WTC}(cC_f)| = |\text{omega}(cC_f)| - 2$.

\begin{enumerate}
\item The smallest $f > 134$ such that
    $|\text{WTC}(cC_{f-1}+1)| = |\text{omega}(cC_{f-1}+1)|$ is $848$.
  The code-length is $1697$ bits.
\item The smallest $f > 134$ such that
    $|\text{WTC}(cC_{f-1}+1)| > |\text{omega}(cC_{f-1}+1)|$ is $3389$.
  The code-lengths are $6779$ and $6778$ bits, respectively.
\item The smallest $f > 134$ such that
    $|\text{WTC}(cC_f)| > |\text{omega}(cC_f)|$ is $13,877,006$.
  The code-lengths are $27,754,013$  and $27,754,012$ bits, respectively.
        (There are larger $f$ where $|\text{WTC}(cC_f)| \le |\text{omega}(cC_f)|$.)
\end{enumerate}

\section{Robustness}
\label{sec:robustness}

If a code is used for storing or transmitting a sequence of integers,
as opposed to calculating entropy, the effect of errors may
be of interest.
The addition of extra error-correcting mechanisms is not covered here.
Consider a sequence of integers, $[N_1, ..., N_k]$,
encoded in each of the codes and imagine the consequences of
a bit being flipped in error.

In the Fibonacci code, switching a `1' to a `0' in the
code-word of $N_j$ causes the value of $N_j$ to be
misread unless it is one of the last two `1's.
In the latter case the end of $N_j$ is not detected correctly and
it absorbs some or all of the code-word of $N_{j+1}$
depending on which bit is flipped and on whether or
not $N_{j+1}$ starts with a `1'.
If a `0' is flipped to a `1' and this is next to
a genuine `1', $N_{j+1}$ is taken to end prematurely and
an extra integer is apparently inserted.
A single bit error affects one or two integers and
may cause an error in indexing (\myfig{fig:FibErrs}). For example:

\begin{figure}[h]
\begin{align*}
    1)\quad & \texttt{100011 011 ...  = 9 2 ... but}\\
    ~ & \texttt{\underbar{0}00011 011 ...  = 8 2 ...}\\
    ~ & \texttt{1000\underbar{0}1 011 ...  = 48 ...}\\
    ~ & \texttt{10001\underbar{0} 011 ...  = 43 ...}\\
    ~ &\texttt{10\underbar{1}011 011 ...  = 12 2 ...}\\
    ~ &\texttt{100\underbar{1}11 011 ...  = 6 4 ... and} \\~\\
    2)\quad & \texttt{100011 1011 ... = 9 4 ... but}\\
    ~ &\texttt{100\underbar{1}11 1011 ... = 4 1 2 ...}
\end{align*}
(Spaces for readability only; flipped bits are underlined.)
\caption{Example errors and Fibonacci}
\label{fig:FibErrs}
\end{figure}

In the Elias omega code, switching a bit of $N_j$'s value
section causes the value to be misread unless it is
the section's leading `1' that becomes `0'.
In the latter case the section is taken to be a length and
parts of one or more following code-words are mistaken as parts of $N_j$.
If a bit in a length section of $N_j$ is flipped,
that length is misread and too little or too much is
taken for the next section unless it is the lead `0' that becomes a `1'.
In that case the length section is taken to be the value section,
$N_j$ is taken to end prematurely, and the rest of $N_j$'s
code-word causes further mistakes.
A single bit error may affect one \textit{or many} integers.

In the Wallace tree code, changing a `1' to a `0' in $N_j$
causes a premature end to the code-word
(not necessarily at the change) and the remainder is
taken as two extra integers.
Changing a `0' to a `1' causes $N_j$ to absorb
$N_{j+1}$ and $N_{j+2}$.
A single bit error affects one or three integers and
causes an error in indexing (fig.\ref{fig:WTCErrs}). For example:

\begin{figure}[h]
\begin{itemize}
\item \texttt{10100 11000 100 ... = 3 4 2 ...} but
\item \texttt{10\underbar{0}00 11000 100 ... = 2 1 1 4 2 ...}
\item \texttt{101\underbar{1}0 11000 100 ... = 90 ...}
\end{itemize}
\caption{Example errors and WTC1}
\label{fig:WTCErrs}
\end{figure}

\section{Asymptotic Analysis of Wallace tree code}
\label{sec:asypmtotics}

It is of interest to determine the asymptotic behaviour of 
WTC for increasing integer $N$. Let 
$L(N)$ denote the length of the binary code-word assigned to 
integer $N$ by WTC1. Recall from Section~\ref{sec:WTC} 
that $C_f$ denotes the $f^{\rm th}$ Catalan number, and $cC_f$ 
denotes the sum of the first $f$ Catalan numbers. Then, the length, 
in bits, of the code-word assigned by WTC1 to 
integer $N$ is
\begin{equation}
	\label{eq:exact:wtc:codelength}
	L(N) = 2 \, f(N) + 1
\end{equation}
where
\[
	f(N) = \inf_{f} \left\{ \mathbb{Z} : N > cC_f \right\}
\]
denotes the smallest integer $f$ such that $N$ exceeds $c C_f$.

\subsection{Bounds on $L(N)$}

The following lemma provides appropriate upper and lower bounds for $L(N)$. \\

\noindent {\bf Lemma 1}. {\em Let $L(N)$ denote the length of the code-word assigned to integer $N$ by WTC defined by (\ref{eq:exact:wtc:codelength}). Let
\[
	\overline{f}(N) \equiv \overline{f} = \frac{ \left( \log N + \frac{3}{2} \log \left( \frac{\log N}{\log 4} \right) + \frac{1}{2} \log \left( \frac{9 \pi}{16} \right) \right)}{ \left(1 - 3/2/\log N\right) \log 4 }
\]
and
\begin{equation}
	\label{eq:n:bar}
	\underline{f}(N) \equiv \underline{f} = \frac{\log N}{\log 4 - (3/2/\overline{f}) \log \overline{f}}.
\end{equation}
Then, for all $N \geq 5$, we have}
\[
    2 \underline{f}(N) + 1 < L(N) < 2 \overline{f}(N)  + 1.
\]

These bounds allow us to determine the asymptotic behaviour of the length of WTC1 code-words. \\

\ignore{
%
}
{
\begin{equation}
	\label{eq:asymp:wt:code}
    \text{\noindent {\bf Theorem 1}.}~ 
	L(N) = \log_2 N + \frac{3}{2} \log_2 \log_2 N + \varepsilon_N,\quad\quad
\end{equation}
where}
\begin{eqnarray}
	\label{eq:asymp:wt:limsup}
	\limsup_{N \to \infty} \{ \varepsilon_N \} &\leq& \frac{3 + \log \left( \frac{9 \pi}{32} \right)}{\log 4} < 2.0748, \\
\label{eq:asymp:wt:liminf}	
	\liminf_{N \to \infty} \{ \varepsilon_N \} &\geq& -\frac{1}{2}.
\end{eqnarray}

The proofs of Lemma 1 and Theorem 1 are deferred to Appendix A.
Complementing the comparisons in section~\ref{sec:comparative}, 
an interesting consequence of Theorem 1 is that there exists 
some integer $M$ such that  
$\forall N>M$, $|\text{WTC}(N)|>|\text{omega}(N)|$, 
although the precise $M$ is unknown 
and likely inconceivably large.

Theorem 1 can be used to demonstrate both the universality and asymptotic optimality of WTC, building on Elias~\cite{Eli75}. To achieve this we first note that the Elias delta code~\cite{Eli75}, which has asymptotic code-length $\log_2 N + 2 \log_2 \log_2 N + O(1)$, is both universal and asymptotically optimal. From Theorem 1 it is clear that code-words of WTC is asymptotically shorter than those of Elias delta code. This establishes the universality and asymptotic optimality of
WTC.

\begin{figure}
\centering
\includegraphics[width=0.3\textwidth,bb=70 20 380 370]{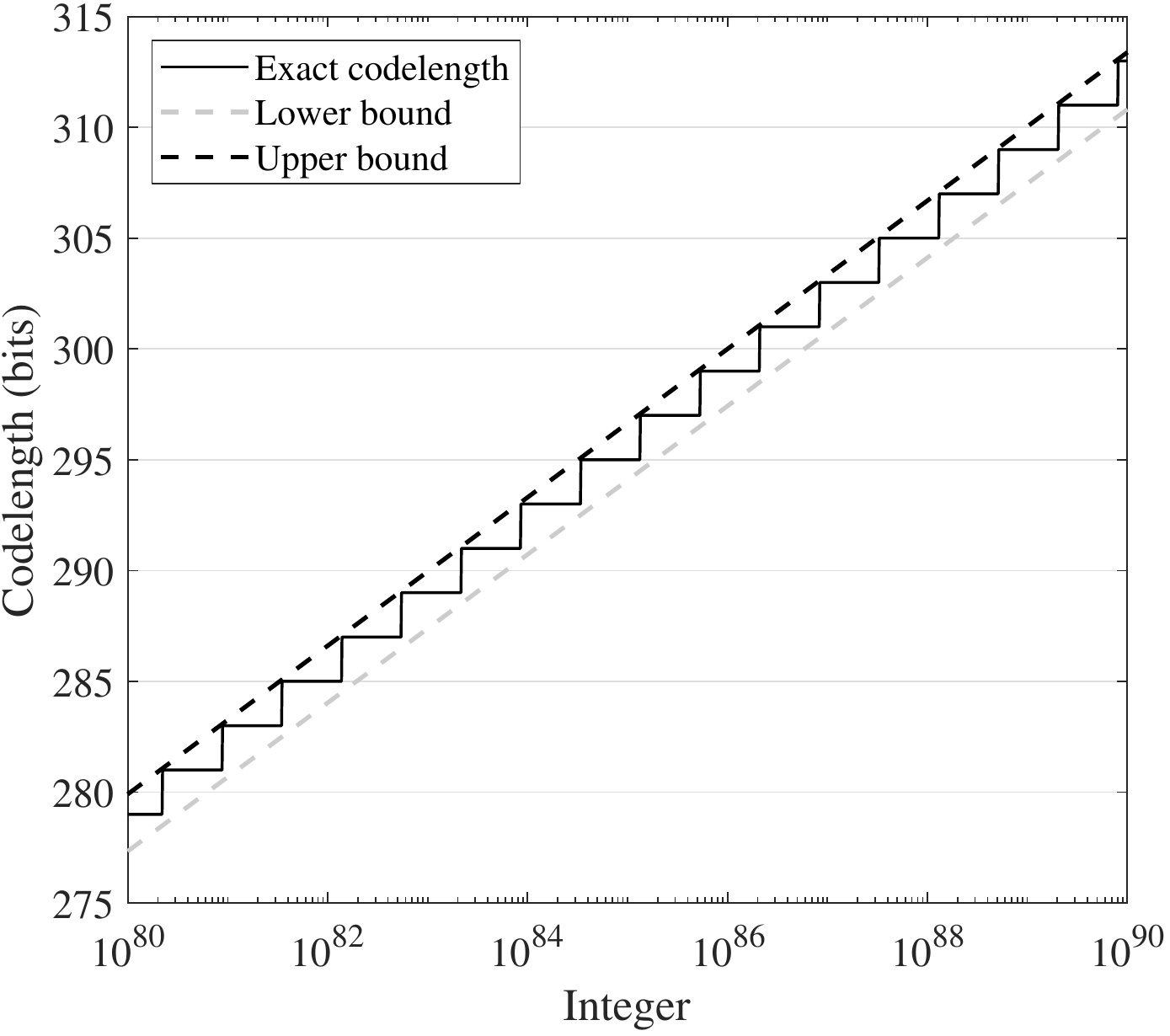}
\caption{Comparison of exact Wallace-tree code-lengths against upper and lower bounds derived in Lemma 1.}
\label{fig:WTC:code:fig}
\end{figure}

\subsection{Asymptotic code-length formulas for WTC}
\label{sec:clformulas}
It is useful to have a simple expression for the code-word lengths of integers under WTC. The requirement for such lengths arises in inductive inference by minimum encoding. It is common to use Rissanen's $\log^*$ code-word length formula to provide an approximate length for the statement of integer parameters. WTC provides an alternative coding scheme in such settings. Theorem 1 suggests the approximate code-word length formula for WTC as:
\begin{equation}
	\label{eq:asymp:wt:code:l}
	L(N;c) = \left\{\begin{array}{ll}
			   		1 & \;\mbox{for } N = 0 \\
			   		3 & \; \mbox{for } N = 1 \\
					\log_2 N + \frac{3}{2} \log_2 \log_2 N + c & \; \mbox{for } N \geq 2
			\end{array} \right.
\end{equation}
where $c$ is a constant. Possible choices for $c$ are:
\begin{itemize}
    \item $c=2$, based on the upper-bound on $\varepsilon_N$ in Theorem 1, which ensures $L(N;c)$ is non-decreasing;
    \item $c=-0.5$, based on the lower-bound on $\varepsilon_N$; or 
    \item $c=0.75$, which is the average of the two error bounds.
\end{itemize}

The accuracy of both the bounds given by Lemma 1 and the asymptotic expression (\ref{eq:asymp:wt:code:l}) is demonstrated in \myfig{fig:WTC:code:fig}. The figure shows a close correspondence between the asymptotic expression (\ref{eq:asymp:wt:code}) and the exact code-length (\ref{eq:exact:wtc:codelength}), particularly as $N$ increases.

\section{Conclusions}
\label{sec:conclusions}

The Wallace tree code (WTC1) for positive integers $N \ge 1$ has
shorter code-words than the Elias omega (and Fibonacci) codes for
most integers upto at least $2.6855 \times 10^{505}$.
Code-word length increases in steps of two from
time to time as $N$ increases.
When using the code to store or transmit a sequence of integers,
the effect of a bit error is localised.
A formula for the approximate code-word length was
derived in section~\ref{sec:clformulas}.

We note that there is a second recursive version of the code, $\text{WTC}_\text{r}$.
It has the same code-word lengths as WTC but
is based on a non-lexicographical ordering of code-words:
For code-words of length $2f+1, f>1$, consider all
partitions of $2f$ into $j$ and $k$ such that $f=j+k$.
Order code-words of the form `1'++$v$++$w$, where
sub-code-words $|v|=j, |w|=k$ and $j+k=2f$, on $v$ and
within that on $w$, recursively.

As discussed in section \ref{sec:variations},
the standard Elias omega code in effect uses a unary code
(``0...'' $\Rightarrow$ length section,
 ``1...'' $\Rightarrow$ final value section)
to indicate the number ($\ge 1$) of sections in a code-word.
This unary code can be replaced by
another code for positive integers, even recursively,
giving the omega\textsuperscript{*} code
which is more efficient for huge values.

Importantly, ordering the various codes discussed on
increasing asymptotic {\em efficiency} gives:
Fibonacci, Elias delta, Wallace WTC, Elias omega,
omega\textsuperscript{2} (sec.\ref{sec:variations}), and
omega\textsuperscript{*}
(all but Fibonacci are asymptotically optimal in the sense of Elias~\cite{Eli75}).

\section*{Acknowledgment}

The authors would like to thank the late Chris Wallace ($1933$--$2004$).

\appendices

\section{Proof of Lemma 1}

The basic approach that we use is to lower and upper-bound the function $f(N)$ with two new functions $\underline{f}(N)$ and $\overline{f}(N)$, respectively. To do this, we find lower and upper-bounds, say $c \underline{C}_f$ and $c\overline{C}_f$ that are continuous in $f$, and solve both $N = c \underline{C}_f$ and $N = c \overline{C}_f$ for $f$. We first derive the upper-bound $\overline{f}$. Our starting point is the following lower-bound on $c C_f$ established by
Topley~\cite{Top16}:
\begin{equation}
	\label{eq:c:C_n:bound:1}
	c C_f > \frac{4^{f+1}}{3 (f+1) \sqrt{\pi f}}.
\end{equation}
Setting the right-hand-side of (\ref{eq:c:C_n:bound:1}) to $N$ and taking logarithms of both sides yields
\begin{equation}
	\label{eq:c:C_n:bound:2}
	(f+1) \log 4 - \frac{1}{2} \log 9 \pi - \log (f+1) - \frac{1}{2} \log f = \log N.
\end{equation}
We wish to solve the above equation for $f$, but a closed form solution does not exist due to the troublesome logarithmic term. Instead, we can use the bounds
\begin{eqnarray*}
	-\log (f+1) - \frac{1}{2} \log f &>& -\frac{3}{2} \log (f+1), \\
	&>& -\frac{3}{2} \log n - \frac{3}{2}, \\
	&>& - \frac{3 f}{2 f_0}  - \frac{3}{2} \log f_0,
\end{eqnarray*}
$\forall f, f_0 > 1$, where the last step is the result of a first order Taylor series expansion of $-(3/2) \log f$ around the point $f_0$, the convexity of $- \log f$ ensuring that the inequality holds. Using this in (\ref{eq:c:C_n:bound:2}) yields the following lower-bound for $c C_f$:
%
\[
	\log c \underline{C}_f = f \log 4 -  \frac{3 f}{2 f_0} - \frac{3}{2} \log f_0 - \frac{1}{2} \log \left( \frac{9 \pi}{16} \right).
\]
%
Solving $c \underline{C}_f = \log N$ for $f$ yields
\[
	\overline{f}(N; f_0) = \frac{\log N + \frac{3}{2} \log f_0 + \frac{1}{2} \log \left( \frac{9 \pi}{16} \right) }{\log 4 - 3/2/f_0}.
\]
The above lower-bound holds for any value of $f_0>1$, but by a judicious choice of $f_0$ it can be tightened. Ignoring terms of order $o(f)$ in (\ref{eq:c:C_n:bound:2}) and solving for $f$ yields an initial guess at $\overline{f}$ of $f_0 = \log N/\log 4$. Using this in $\overline{f}(N; f_0)$ yields (\ref{eq:n:bar}),
%
%
which satisfies $\overline{f} \geq f(N)$ for all $N \geq 5$. Substituting $\overline{f}$ for $f(N)$ in (\ref{eq:exact:wtc:codelength}) yields the upper-bound.

We now derive $\underline{f}$. We start with the following upper-bound on $c C_f$
\begin{equation}
	\label{eq:c:C_f:bound:4}
	c C_f < \frac{4^{f+1}}{3 \sqrt{\pi f^3} }.
\end{equation}
Setting the right-hand-side of (\ref{eq:c:C_f:bound:4}) equal to $N$ and taking logarithms of both sides yields
\begin{equation}
	\label{eq:c:C_f:bound:5}
	(f+1) \log 4 - \frac{1}{2} \log 9 \pi - \frac{3}{2} \log f = \log N.
\end{equation}
As before, solving (\ref{eq:c:C_f:bound:5}) directly for $f$ is impossible due to the logarithmic term. Instead we note that we can use the bounds
\[
	-\frac{3}{2} \log f < - \left( \frac{3 \log f_1}{2 f_1} \right) {\rm min}\{f, f_1\}
\]
for $f > 1$ and $f_1 > 1$, along with the fact that $\log 4 - (1/2) \log 9 \pi < 0$ to derive the following upper-bound for $c C_f$:
\[
	\log c \overline{C}_f = f \log 4 - \left( \frac{3 \log f_1}{2 f_1} \right) {\rm min}\{f, f_1\}.
\]
We now note due to the strictly increasing nature of $c \overline{C}_f$ that if $f_1 > f(N)$ then the solution $\underline{f}$ of $\log c \overline{C}_f = \log N$ will satisfy $\underline{f} < f_1$. We therefore choose $f_1 = \overline{f}$, which we previously established is an upper-bound to $f(N)$ for $N\geq 5$, and solve for $\underline{f}$, yielding
\[
	\underline{f}(N) \equiv \underline{f} = \frac{ \log N }{\log 4 - (3/2/\overline{f}) \log \overline{f} }
\]
which satisfies $\underline{f} < f(N)$ for all $N \geq 5$. Substituting $\underline{f}(N)$ for $f(N)$ in (\ref{eq:exact:wtc:codelength}) completes the proof. $\hfill$ $\Box$

\section{Proof of Theorem 1}

Let
\[
	L_A(N) = \log_2 N + \frac{3}{2} \log_2 \log_2 N,
\]
and let $\varepsilon_N = L(N) - L_A(N)$. If we rewrite $\overline{f}$ as
\[
	\overline{f} = \frac{ L_A(N) + \frac{1}{2} \log_2 \left( \frac{ 9 \pi}{128} \right)}{ 2 \left(1 - 3/2/\log N\right) }
\]
then a straightforward application of L'Hopital's rule shows that
\[
	\lim_{N \to \infty} \left\{ 2 \overline{f} + 1 - L_A(N) \right\} = \frac{3 + \log \left( \frac{9 \pi}{32} \right)}{\log 4},
\]
which itself implies (\ref{eq:asymp:wt:limsup}) if we note that $\varepsilon_N \leq 2 \overline{f} + 1 - L_A(N)$ (by application of Lemma 1). Similarly, by tedious algebra we can show that
\[
	\lim_{N \to \infty} \left\{ 2 \underline{f} + 1 - L_A(N) \right\} = -\frac{1}{2},
\]
which implies (\ref{eq:asymp:wt:liminf}) as $\varepsilon_N \geq 2 \underline{f} +1 - L_A(N)$, completing the proof. $\hfill$ $\Box$

%
%



\ifCLASSOPTIONcaptionsoff
  \newpage
\fi



\bibliographystyle{IEEEtran}

\bibliography{paper}
\end{document}